\documentclass[review]{elsarticle}

\usepackage{lineno,hyperref}
\usepackage{graphicx}
\usepackage{bm}
\modulolinenumbers[5]

\begin{document}

\begin{frontmatter}

\title{A Multi-length Bunches Design for Electron Storage Rings with Odd Buckets}

\author{Liangjing Zhu}
\address{National Synchrotron Radation Laboratory, University of Science and Technology of China, Hefei ,230026, China}
\address{SLAC National Accelerator Laboratory, Menlo Park, California 94025, USA}

\author{Dao Xiang}
\address{Key Laboratory for Laser Plasmas (Ministry of Education),
Department of Physics and Astronomy, Shanghai Jiao Tong University, Shanghai
200240, China}

\author{Lin Wang,Weimin Li\footnote{E-mail address:lwm@ustc.edu.cn \\ Phone number:+86055163602105 \\ Submitted to  Chinese Physics 
C}}
\address{National Synchrotron Radation Laboratory, University of Science and Technology of China, Hefei ,230026, China}

\author{Xiaobiao Huang}
\address{SLAC National Accelerator Laboratory, Menlo Park, California 94025, USA}

\begin{abstract}
A scheme with two superconducting RF cavities (sc-cavities) is designed to upgrade electron storage rings with odd buckets into multi-length bunches. In this paper, Hefei Light Source II (HLS II) is given as an example for odd buckets. In accordance with 45 buckets, which is multiples of 3, three different length of bunches generated simultaneously is proposed in the presently applied user optics. The final result is to, without low-$\alpha$ optics, fill HLS II with long bunches of 50 ps length, medium bunches of 23 ps and short bunches of 6 ps. Each third buckets can be filled with short bunches, of which the current limit is up to 6.6 mA, more than 60 times the value of low-$\alpha$ mode. Moreover, particles tracking about beam dynamics performed by ELEGANT and calculations about beam instabilities are presented in this paper.
\end{abstract}


\end{frontmatter}

\linenumbers

\section{Introduction}
An increasing interest in short x-ray pulses requires short electron bunches in storage rings. Over the years, the traditional way to get short bunches is to decrease the momentum compaction factor by low-$\alpha$ optics \cite{MLS}. However, the average current per bunch will decrease to the order of $\mu$A magnitude with low-$\alpha$, limited by beam instability and microbunching due to coherent synchrotron radiation and other collective effects \cite{stupakovBeaminstablity}. 

BESSY II presented an idea to produce long and short bunches alternatively in their storage ring: with two superconducting cavities, 3rd harmonic and 3.5th harmonic of the fundamental RF cavity, the voltage gradient produced by two sc-cavities could add up at even points for bunches focusing to get short bunches, and cancel each other at odd points to get long bunches \cite{BessyIIupgrade}. The new method greatly improved the capacity of storing current in the ring. Nonetheless, BESSY II is filled with 400 buckets, which is the multiples of 2, leading to the result of choose the second sc-cavity as a 1/2 times higher harmonic one. What if the ring is with odd buckets, such as multiples of 3, 5, 7? In this paper,  we find the problem could be solved by choosing the frequency of the second sc-cavity (sub harmonic cavity) in a different way. For a ring, whose minimum common factor is $k$ ($k>1$), the frequency of the sub harmonic cavity could be chosen $1/k$ times higher  than harmonic one. Several typical situations are listed in Table.\ref{tab:cavity harmonic number} , in which $N_{i} (i=1,2,3,4)$ is chosen accord to the  voltage and frequency of  original cavity in different rings.

\begin{table}[htbp]
  \centering
  \caption{\label{tab:cavity harmonic number}Harmonic Number with Different Buckets Number}
 \begin{tabular}{ccc}
 \hline
  Common factor  & Harmonic & Sub harmonic \\
   2& $N_{1}$ &$N_{1}+1/2$ \\
   3& $N_{2}$ &$N_{2}+1/3$  \\
   5& $N_{3}$ &$N_{3}+1/5$  \\
   7& $N_{4}$ &$N_{4}+1/7$ \\
    \hline
  \end{tabular}
\end{table}

In addition, multi-cell structure is required for BESSY II and also other rings to achieve the 20 MV voltage of new sc-cavities because of the high original cavity voltage, and multi-cell structure may lead to some HOM problems. So, if we could find a ring with low voltage, the HOM problems may be much simpler. 

In this paper, we present a new scheme for HLS II, it is a good example to solve the odd buckets problem, and make the HOM problems easy at the same time.  An idea with two sc-cavities whose voltages are around 2 MV and frequencies are around 1.3 GHz, is designed for HLS II. A 1/3 times higher harmonic sc-cavity is expected to be applied in HLS II, and with careful  phase tuning,  medium, long and short bunches are generated simultaneously. Moreover, because of the low voltage and frequency of fundamental cavity,  it is easy to increase the voltage gradient in the ring by 100 times, and single-cell or two-cell structure is enough for two sc-cavities.

\section{The Lattice of HLS II Storage Ring}
\begin{figure}[htbp]
  \centering
  \includegraphics[width=0.68\textwidth]{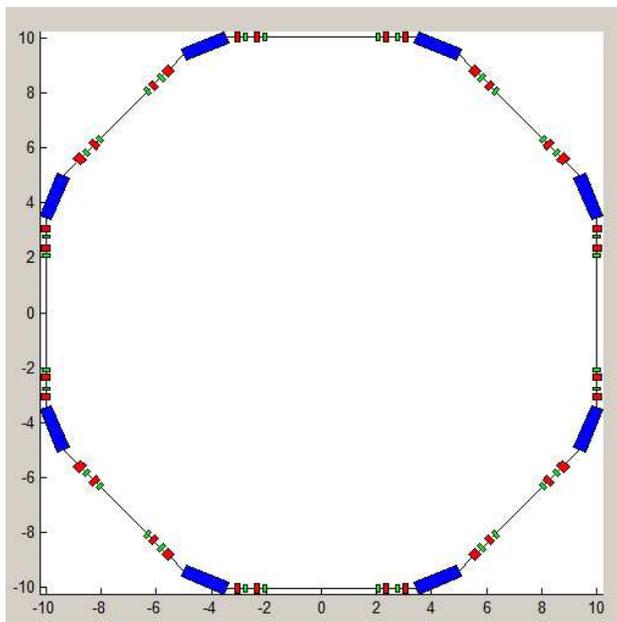}\\
  \caption{\label{fig: HLS magnet} Magnet layout of HLS II. Bends are in blue rectangles, quadrupoles are in red rectangles and sextupoles are in green rectangles.The HLSII ring is a typical octagonal structure.}
\end{figure}

\begin{figure}[htbp]
  \centering
   \includegraphics[width=0.9\textwidth]{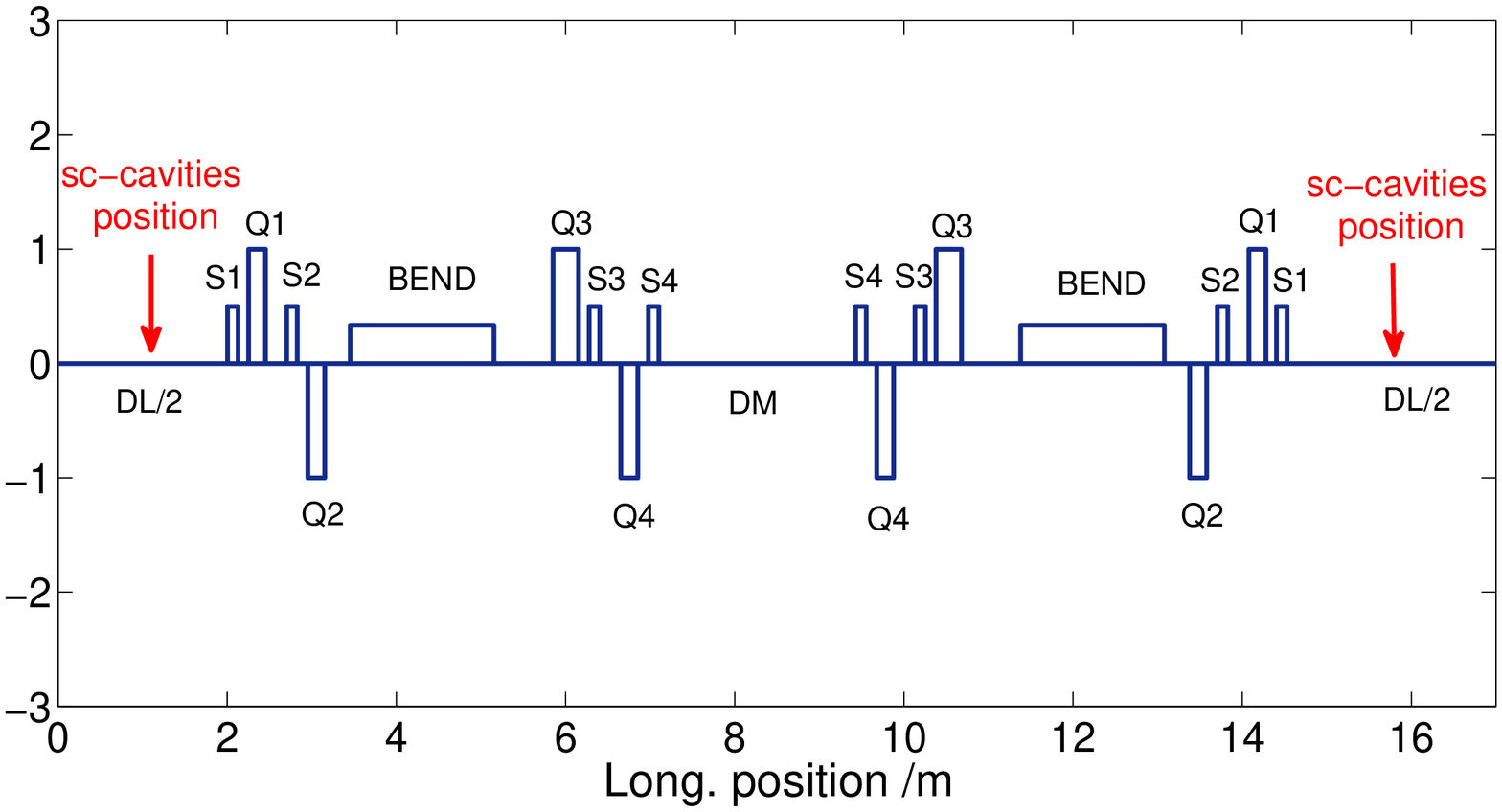}
  \caption{\label{fig: HLS lattice} The lattice structure per cell in HLS II}
  \includegraphics[width=0.9\textwidth]{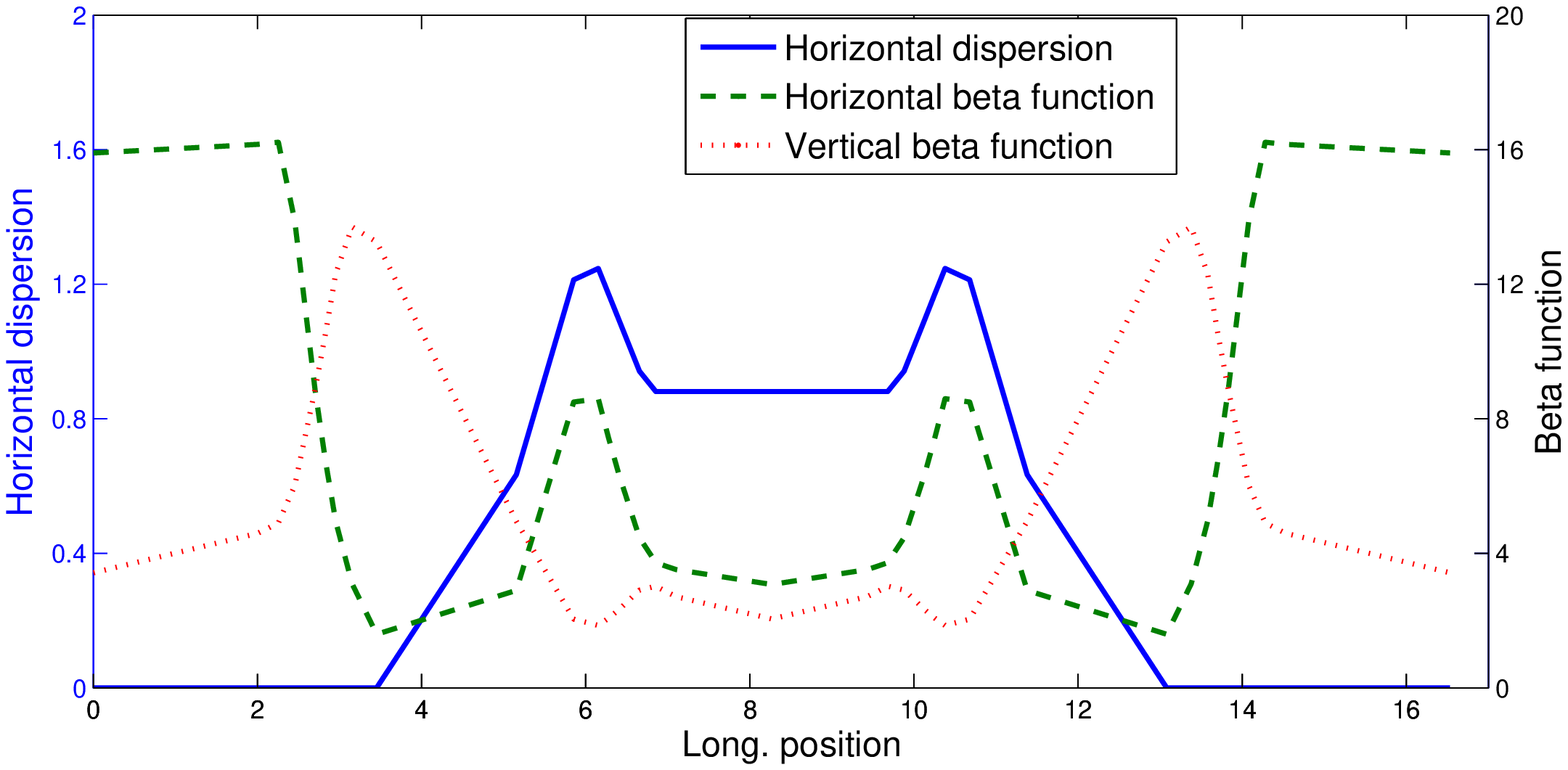}
  \caption{\label{fig: beta function} $\beta$ and dispersion function of HLS II. The new superconducting cavities are required to be located at the zero dispersion position shown in Figure 2.}
\end{figure}

Hefei Light Source II \cite{HLSupgrade}is an 800 Mev electron storage ring, whose main parameters are given in Table.\ref{tab: general para}. The lattice structure of HLS II was chosen as a DBA with 4 periods, and 8 straight sections which include four 4.0 m long straight sections and four 2.3 m short chromatic straight sections.  In each half cell, there are four quadrupoles and four combined function sextupoles. The magnet layout of ring is shown in Fig.\ref{fig: HLS magnet}  and Fig.\ref{fig: HLS lattice}.

\begin{table}[htbp]
  \centering
  \caption{\label{tab: general para} General Parameters of HLS II.}
 \begin{tabular}{l l}
 \hline
    Nominal energy & 800 MeV \\
    Circumference & 66.13 m \\
    Number of cells & 4 \\
    Number of dipoles & 8 \\
    Bending radius & 2.1645 m \\
    RF frequency & 0.204 GHz \\
    Energy spread & 0.00047 \\
    Emittance & 36.38 nm $\cdot$ rad \\
    Beam current & $>$ 300 mA \\
    Momentum compaction & 0.0205 \\
    Damping time ($\tau_{x},\tau_{y},\tau_{s}$) & (20.00, 21.08, 10.84) ms \\
    Tunes($\nu_{x},\nu_{y}$) & (4.44, 2.80) \\
    Nature chromaticity($\xi_{x},\xi_{y}$) & (-9.89, -4.66) \\
    Energy loss per turn & 16.74 keV \\
    \hline
  \end{tabular}
\end{table}

There are two operation modes in HLS II, Mode A is an achromatic mode whose dispersion in the long straight sections is zero, and Mode B is a distribute dispersion mode whose emittance is smaller than Mode A. In this paper, we choose the achromatic mode (Mode A) to make simulation. The main parameters of HLS II are summerized in Table.\ref{tab:lattice para}, and Fig.\ref{fig: beta function} shows the $\beta$ and dispersion function per cell.

\begin{table}[htbp]
  \centering
  \caption{\label{tab:lattice para} The Main parameters of Lattice(Half cell)}
\begin{tabular}{l l}
\hline
Start point & Midpoint of DL \\
 DL/2  & 2.003175 m \\
 S1      & 0.00 m$^{-3}$ \\
 Q1     & 3.8807 m$^{-2}$ \\
 S2     & 0.00 m$^{-3}$ \\
 Q2     & -3.2031 m$^{-2}$ \\
 DBQ1  & 0.30 m \\
 B         & 1.7 m / 1.2336 T \\
 DBQ2  & 0.70 m \\
 Q3      & 3.7871 m$^{-2}$\\
 S3      & 49.36 m$^{-3}$ \\
 Q4     &  -3.3874 m$^{-2}$ \\
 S4      & -79.07 m$^{-3}$ \\
 DM/2  & 1.163175 m \\
 Symmetric point & Midpoint of DL\\
 \hline
 \end{tabular}
 \end{table}
 
\section{The Medium-Long-Short Bunches Scheme}

In an electron storage ring, the equilibrium bunch length is \cite{wiedemann2003particle}
 \begin{equation}
    \sigma_{s} = \frac{c\,\delta_{\varepsilon}}{2\pi} \sqrt{\frac{4{\pi}^{2} \alpha E_{0}}{ce f_{rev} V' }} 
 \end{equation}
here $f_{rev}$ is the revolution frequency, $\alpha$ is the momentum compaction factor, $V'$ is the voltage gradient, $\delta_{\varepsilon}$ is the equilibrium energy spread. From this relation, the  bunch length can be shorten by decreasing momentum compaction factor or increasing the voltage gradient for$\sigma_{s} \propto{\sqrt{\alpha /V'}}$. In this section, we just focus on the method of increasing the voltage gradient.

The original cavity is rf-frequency $f_{0}$ = 0.204 GHz, voltage $V_{0}$ = 0.25 MV and phase $ \psi_{s0}$ =3.0747 rad. It is used to replenish the energy loss by synchrotron radiation, and the frequency also leads to the fill pattern of 45 buckets. Because of the high gradient for bunch focusing, two superconducting cavities are required here. The first one is a sixth harmonic cavity $f_{1}$ = 1.2240 GHz ($f_{1}=6 f_{0}$) , and considering the 45 buckets, the second cavity is chosen as ($f_{2}=(6+\frac{1}{3}) f_{0}$), a 1/3 times higher harmonic in rf-frequency. The two sc-cavities can devote to three kinds of bunches.

In addition to the rf-frequency, finite voltage and phase of sc-cavities must be chosen for particle acceleration and stable buckets. The sum voltage gradient at longitudinal distance is
\begin{equation}
 V'_{\texttt{sum}}=\frac{2 \pi}{c} \cdot [fv]_{\texttt{focusing}}
 \end{equation}
 and  $[fv]_{\texttt{focusing}}$ is defined as a rf-focusing parameter which directly dominates the bunch length

\begin{eqnarray}
 [fv]_{\texttt{focusing}}= \Big[f_{0} V_{0} \cos  \Big(\frac{2 \pi f_{0}}{c} s+\psi_{s0}\Big)
 +f_{1} V_{1} \cos  \Big(\frac{2 \pi f_{1}}{c} s+\psi_{s1}\Big)
 \nonumber
 \\
+ f_{2} V_{2} \cos  \Big(\frac{2 \pi f_{2}}{c} s+\psi_{s2}\Big)\Big]
\end{eqnarray}

here $\psi_{s0}$,$\psi_{s1}$,$\psi_{s2}$ is the phase of original, harmonic and sub harmonic cavity. For a proper canceling to get long bunches, frequencies and voltage amplitudes are required to be $f_{1}V_{1}=f_{2}V_{2}$ \cite{BessyIIupgrade}.  Although higher voltage could shorten bunches more, taking the HOM problem and cavities design into account,   voltages are chosen as $V_{1}$=1.9 MV  and  $V_{2}$=1.8 MV.

To get $\psi_{s1}$ and $ \psi_{s2}$, 45 buckets are divided into three groups: Bucket$_{(3m)}$, Bucket$_{(3m+1)}$, Bucket$_{(3m+2)}$,($m = 0,1,2,3,\ldots,14$).  Two conditions are made here to determined the phase for cavities. First, only original cavity is used for energy recovery because of the expensive cooling system for superconducting cavities. Second, voltage gradient produced by two cavities are expected to cancel each other at the position of  Bucket$_{(3m+1)}$ to get long bunches.  Two conditions can be summarized as follows
\begin{equation}
 \begin{array}{l l}
 \texttt{condition 1} & V_{1}\sin(\psi_{s1})+ V_{2}\sin(\psi_{s2})=0\\
 \texttt{condition 2} & \psi_{s1}+\psi_{s2}=\frac{1}{3}\pi+k\pi 
 \end{array}
\end{equation}
here $k$ is an integer.

By solving the conditions, phases are chosen as $\psi_{s1}=2.1412$ rad and $\psi_{s2}=-1.0940$ rad. The main parameters of three cavities are given in Table.\ref{tab:sc-cavities para}.  And two sc-cavities positions require zero dispersion to avoid bunch lengthening by coupling effects \cite{RFcavitiesbook}, shown in Fig.\ref{fig: HLS lattice}.
 \begin{table}[htbp]
  \centering
  \caption{\label{tab:sc-cavities para} The Main Parameters of Three Cavities}
\begin{tabular}{lllll}
\hline
Cavities & Harmonic  &  Frequency & Voltage  & Phase \\
& Number & $f$ (GHz) & $V$ (MV)& $\psi_{s}$  \\
\hline
Original & 1 & 0.204 & 0.25 &  176.17$^\circ$ \\

Harmonic & 6 & 1.2240 &1.9 & 122.68$^\circ$ \\

Sub Harmonic &6$\frac{1}{3}$& 1.2920 & 1.8 & -62.68$^\circ$\\
\hline
\end{tabular}
\end{table}

 After phase tuning, the voltage gradient produced by the two sc-cavities cancel each other at the Bucket$_{(3m+1)}$ to get long bunches, add up at the Bucket$_{(3m+2)}$ to get short ones, and Bucket$_{(3m)}$  can be filled with medium bunches.
 
\begin{figure}
  \centering
  \includegraphics[width=0.9\textwidth]{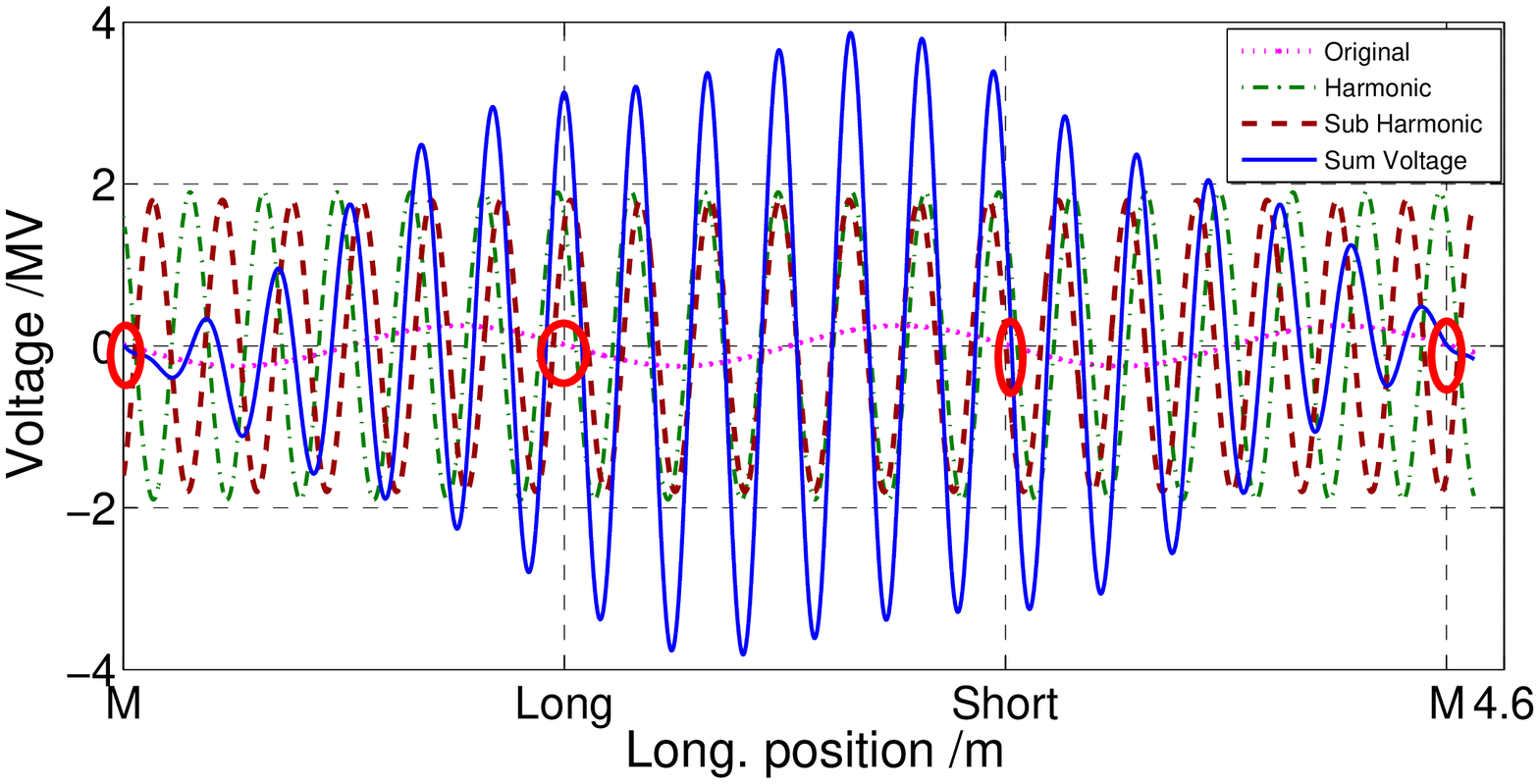}
  \caption{\label{fig:sum voltage} Voltage of Three Cavities.The sum voltage in MV as a function of the longitudinal distance in m is shown in blue. Voltages of the original, harmonic and sub harmonic cavities are shown in purple, green and red.}
   \includegraphics[width=0.9\textwidth]{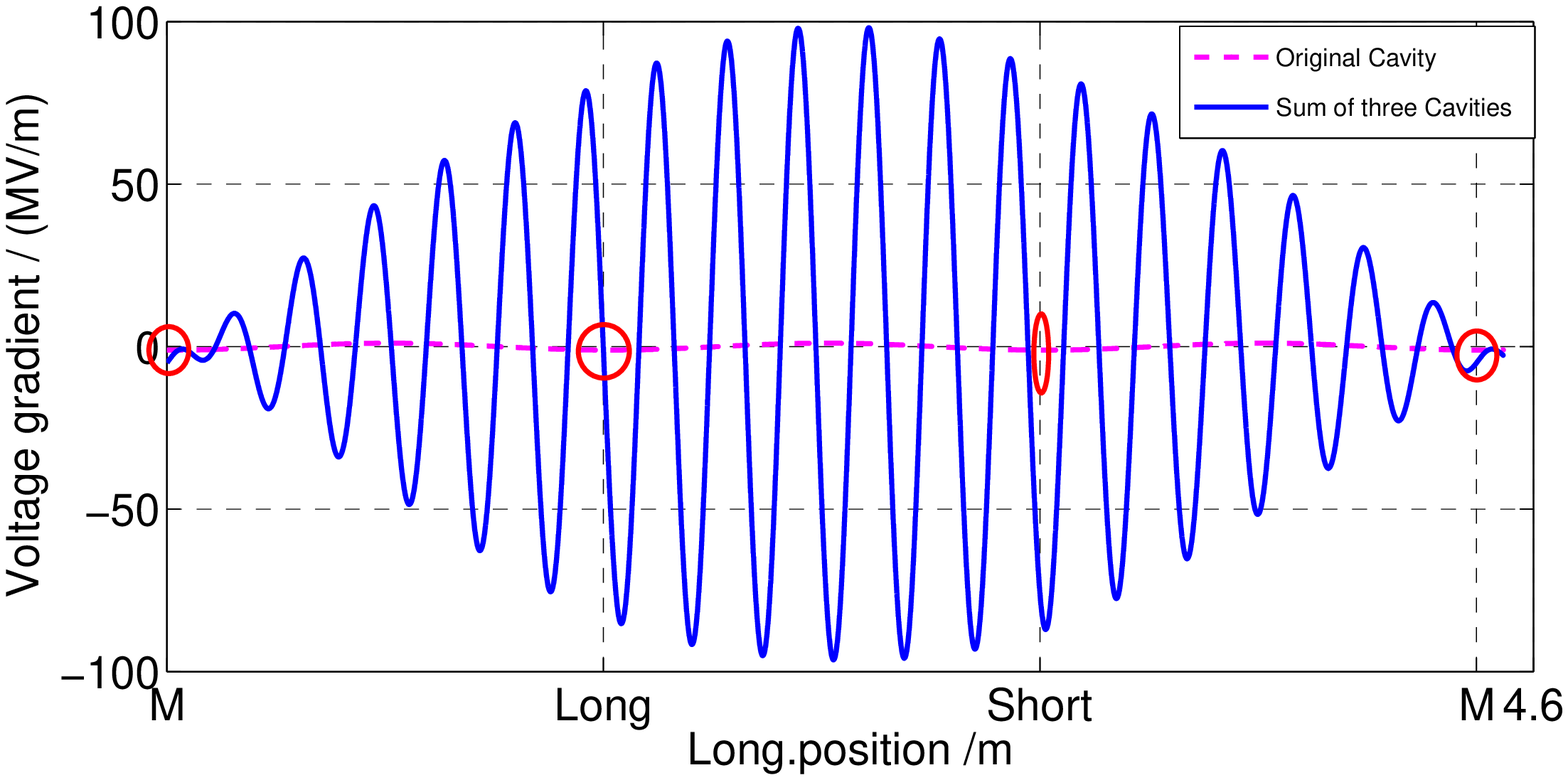}
  \caption{\label{fig:sum voltage gradinet}The Sum Voltage Gradient of Three Cavities. The sum voltage gradient in MV/m as a function of the longitudinal distance in m is shown in blue.The voltage gradient of the original cavity is shown in purple.}
\end{figure}

Fig.\ref{fig:sum voltage} shows the sum voltage of three cavities as a function of the longitudinal position. As $\sigma_{s} \propto {\sqrt{\alpha / V'}}$, the sum voltage gradient as a function of longitudinal position is shown in Fig.\ref{fig:sum voltage gradinet}.  From Fig.\ref{fig:sum voltage gradinet}, it is clearly to see that the voltage gradient increase a lot by two sc-cavities, and three buckets are in one period. The first bucket is filled with a medium bunch, second is long , third is short, and following buckets are just in the same pattern,which are indicated by the red circles in Fig.\ref{fig:sum voltage} and Fig.\ref{fig:sum voltage gradinet}. 

\begin{figure}
  \centering
  \includegraphics[width=0.9\textwidth]{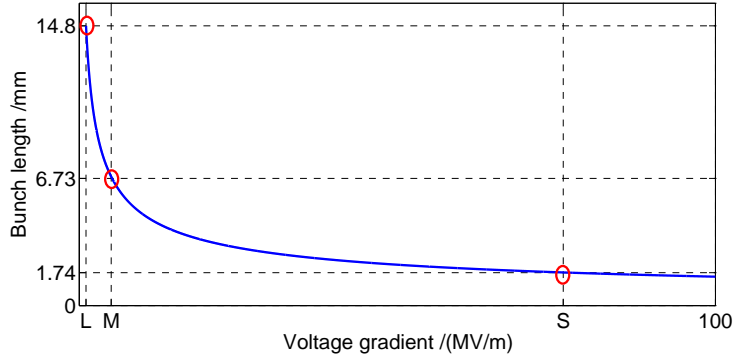}
  \caption{\label{fig:bunch length of different gradient} The bunch length in mm as a function of voltage gradient in MV/m. Three different bunches are indicated in red circles L,M,S for long , medium and short ones.}
\end{figure}

Actually, by applying sc-cavities in HLS II, the number of buckets increases 6 times (to 270 buckets). However, we just care about bunches at the position of original buckets. Long bunches are placed at $3m+1$ multiples of 1.4696 m, short bunches are at $3m+2$ multiples of 1.4696 m, and bunches at $3m$ multiples of 1.4696 m are about 2.2 times shorter than long bunches length. Without Low-$\alpha$ mode, we can get long bunches at 14.8 mm (50 ps), short bunches at 1.74 mm (6 ps) and medium bunches at 6.73 mm (23 ps). The relation between bunch length and voltage gradient are show in Fig.\ref{fig:bunch length of different gradient}.

\section{Dynamic Aperture}
Four different groups of sextupoles per cell shown in Figure 1 are used to correct chromaticity and harmonics in HLS II. In Mode A, the operation mode in this paper, only the sextupoles in short sections are useful for chromaticity corrections because sextuploes must be placed at locations where the dispersion function does not vanish, $\eta_{x}\neq 0$ \cite{wiedemann2003particle}.

\begin{figure}[htbp]
  \centering
  \includegraphics[width=0.9\textwidth]{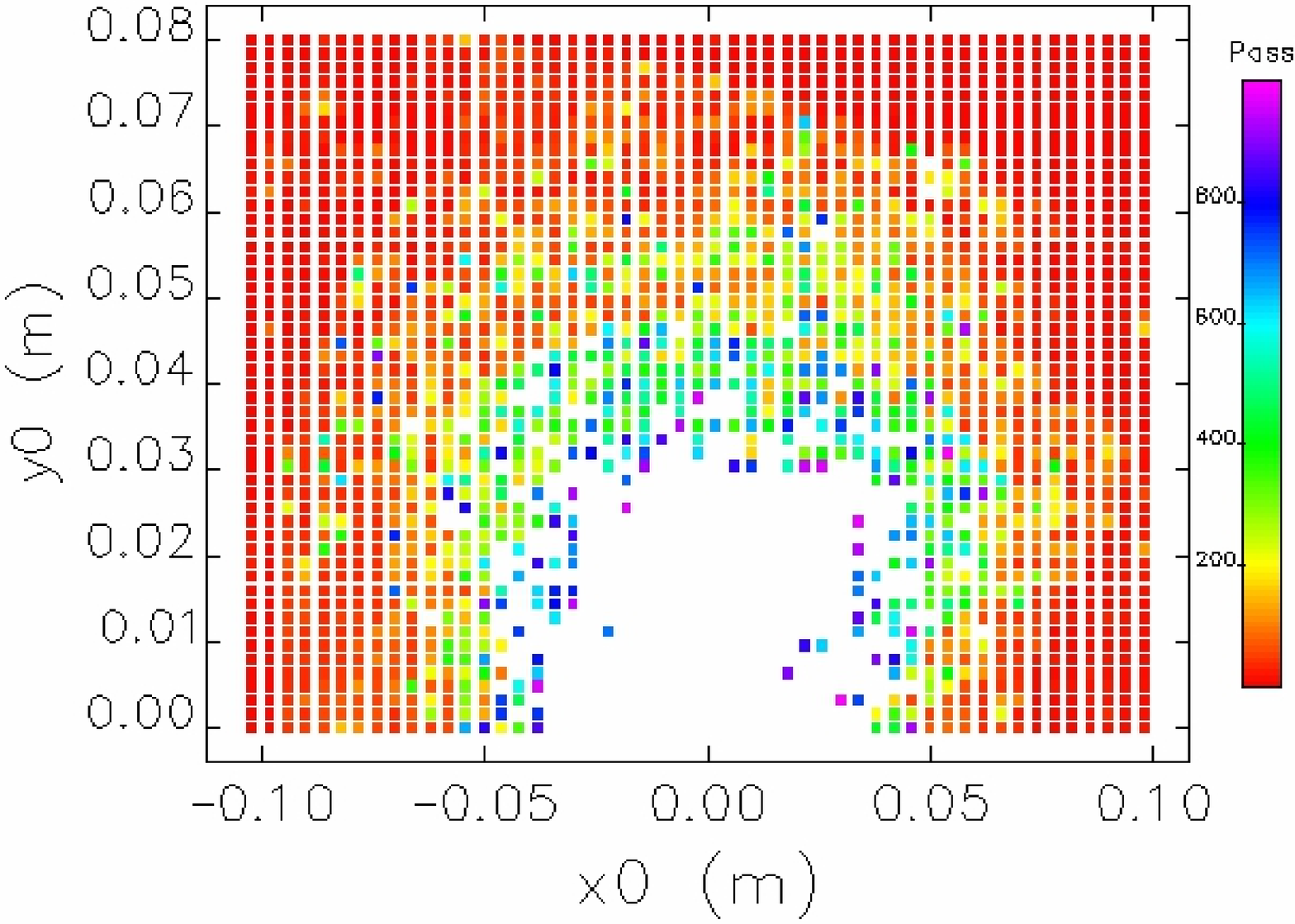}\\
  \includegraphics[width=0.9\textwidth]{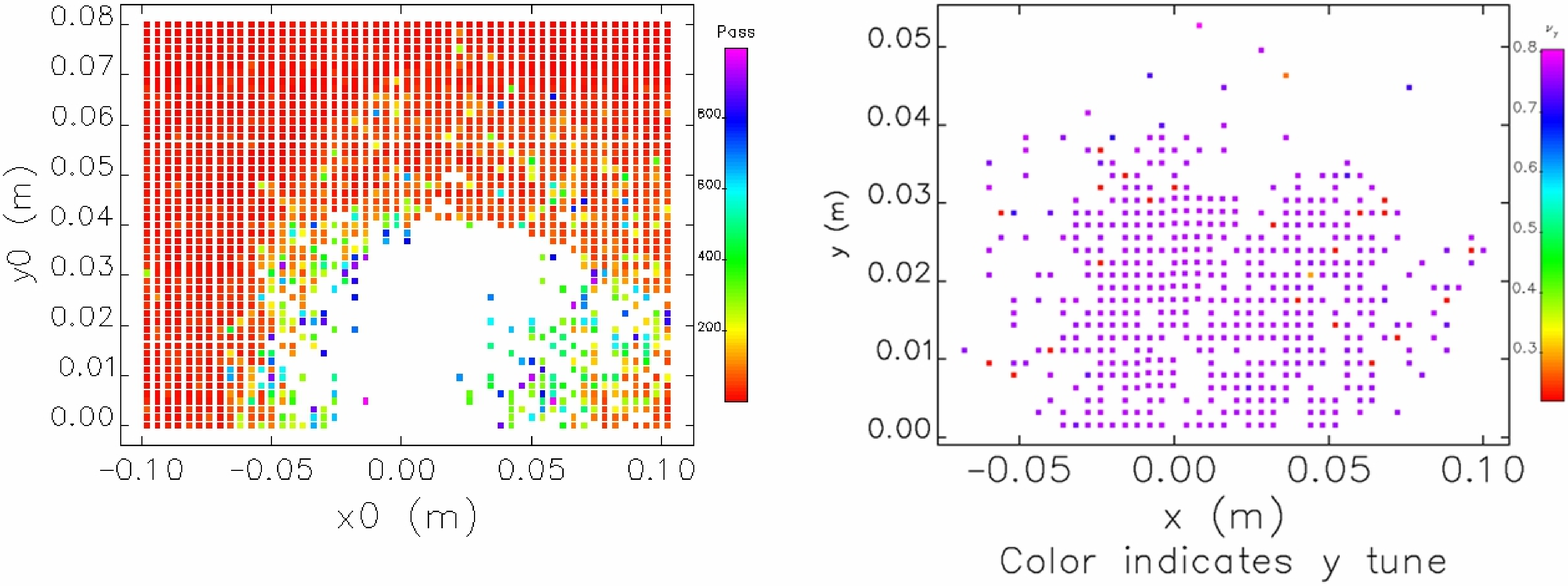}\\
  \includegraphics[width=0.9\textwidth]{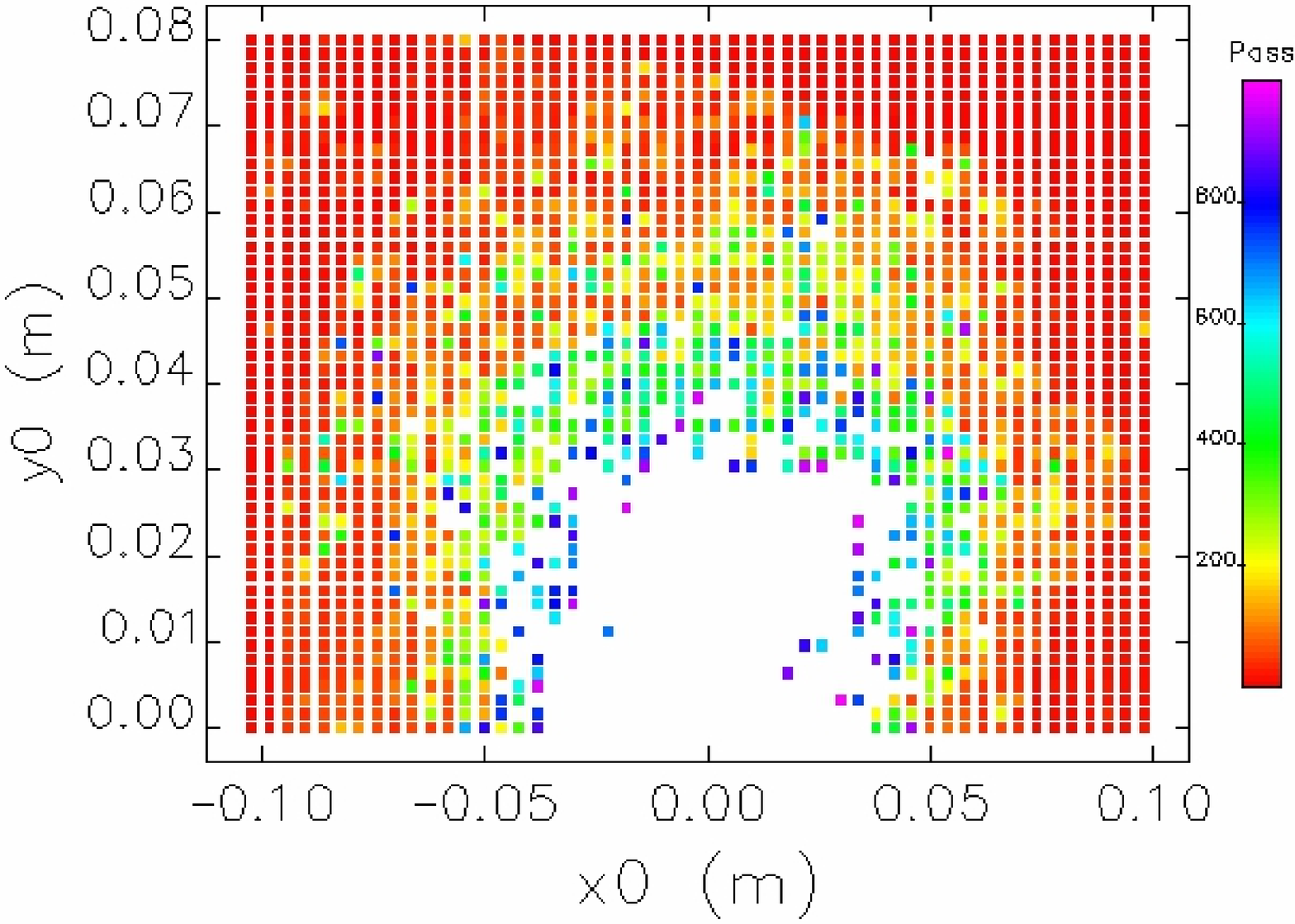}\\
  \caption{\label{fig: dynamic aperture} Dynamic aperture and frequency map analysis for $\delta=0,\pm2 \%$  }
\end{figure}

Simulation for dynamic aperture by tracking 1000 turns is performed by ElegantRingAnalysis \cite{elegantringanalysis}. Fig\ref{fig: dynamic aperture} shows dynamic aperture (DA) and frequency map for different energy spread at $\delta=0,\pm 2 \%$ , which prove that DA in horizontal plane can reach about 40 mm. The aperture is large enough to ensure no particles loss.

\section{Tracking results}
\begin{figure}[htbp]
  \centering
     \includegraphics[width=0.45\textwidth]{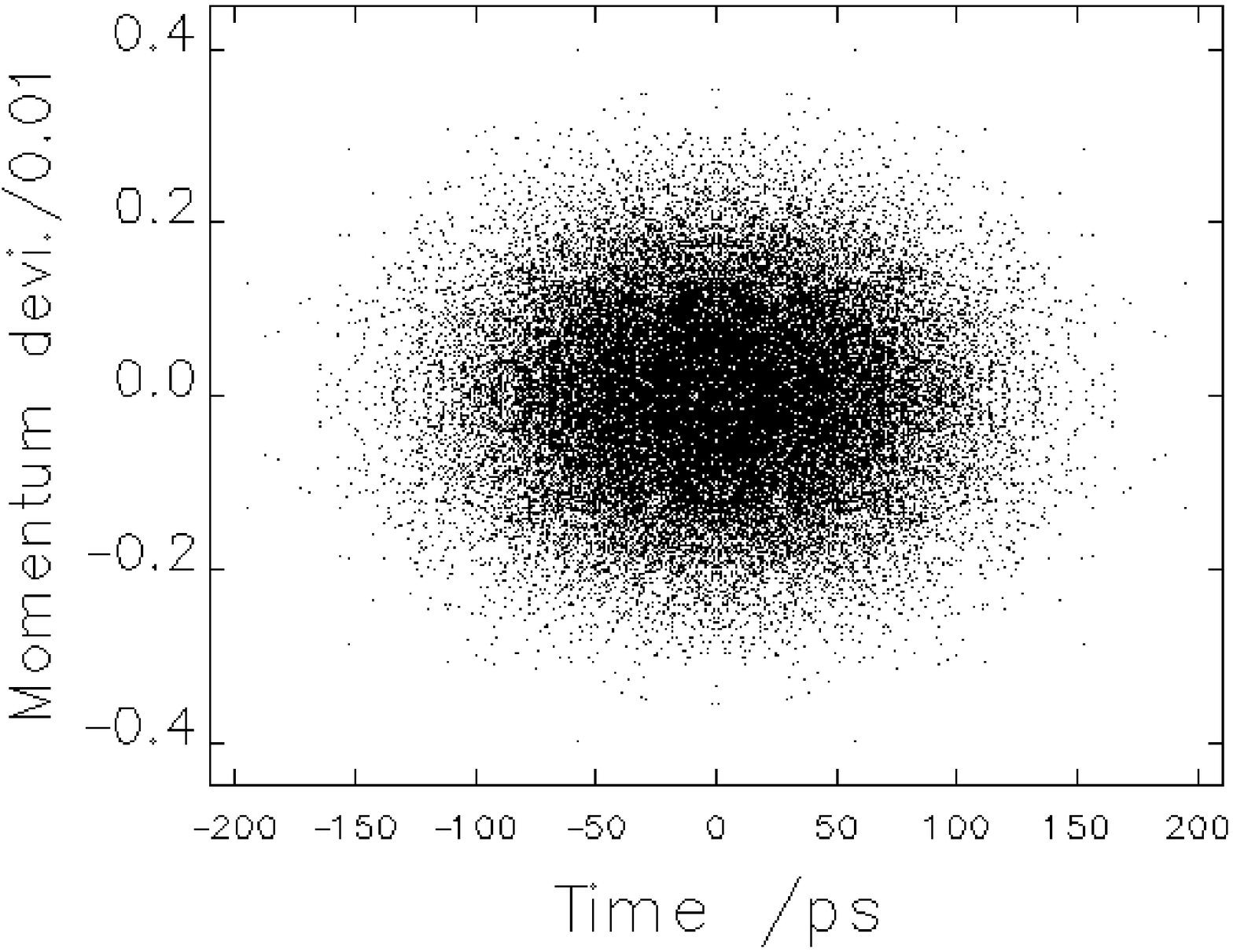} \\
     \includegraphics[width=0.45\textwidth]{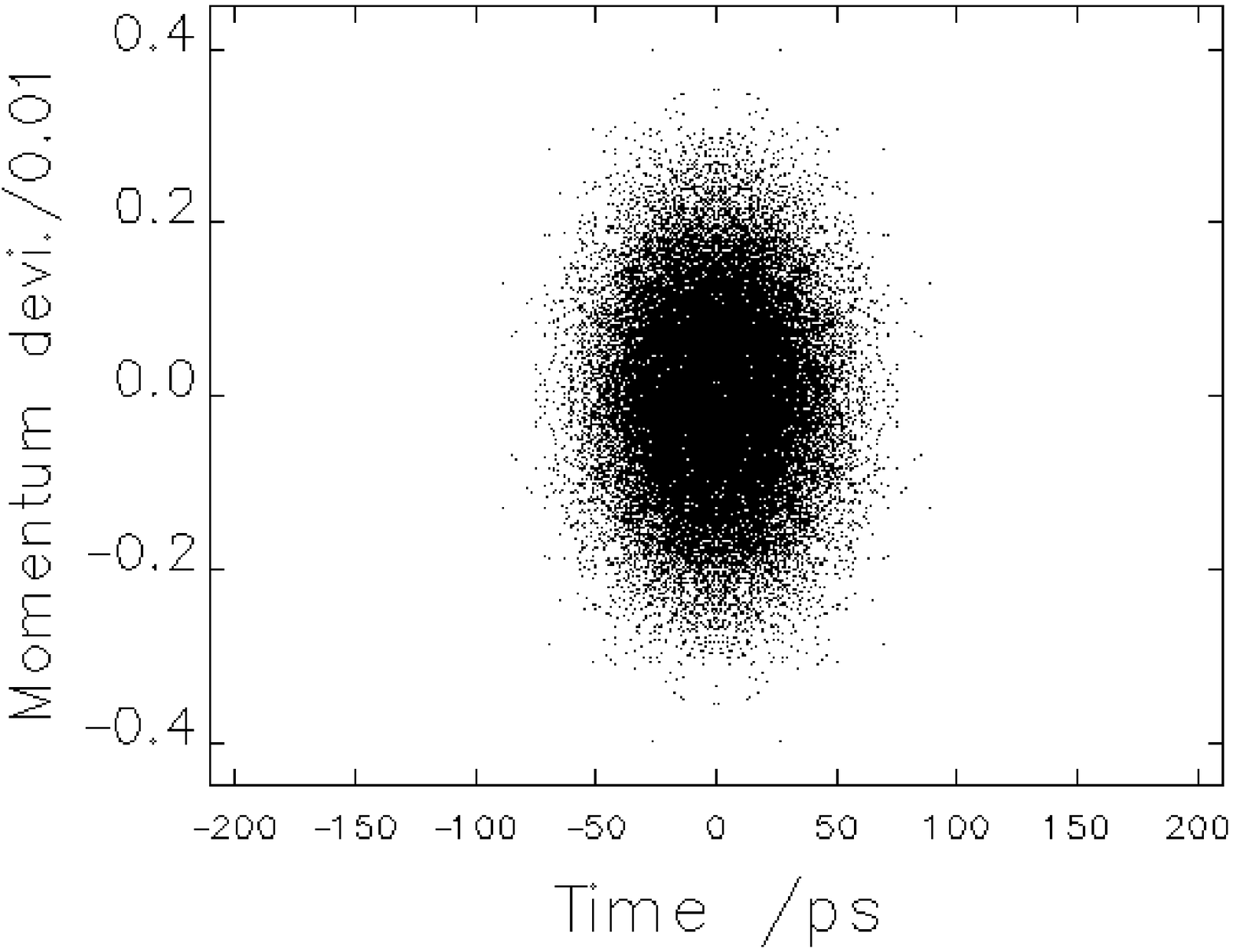}
      \includegraphics[width=0.45\textwidth]{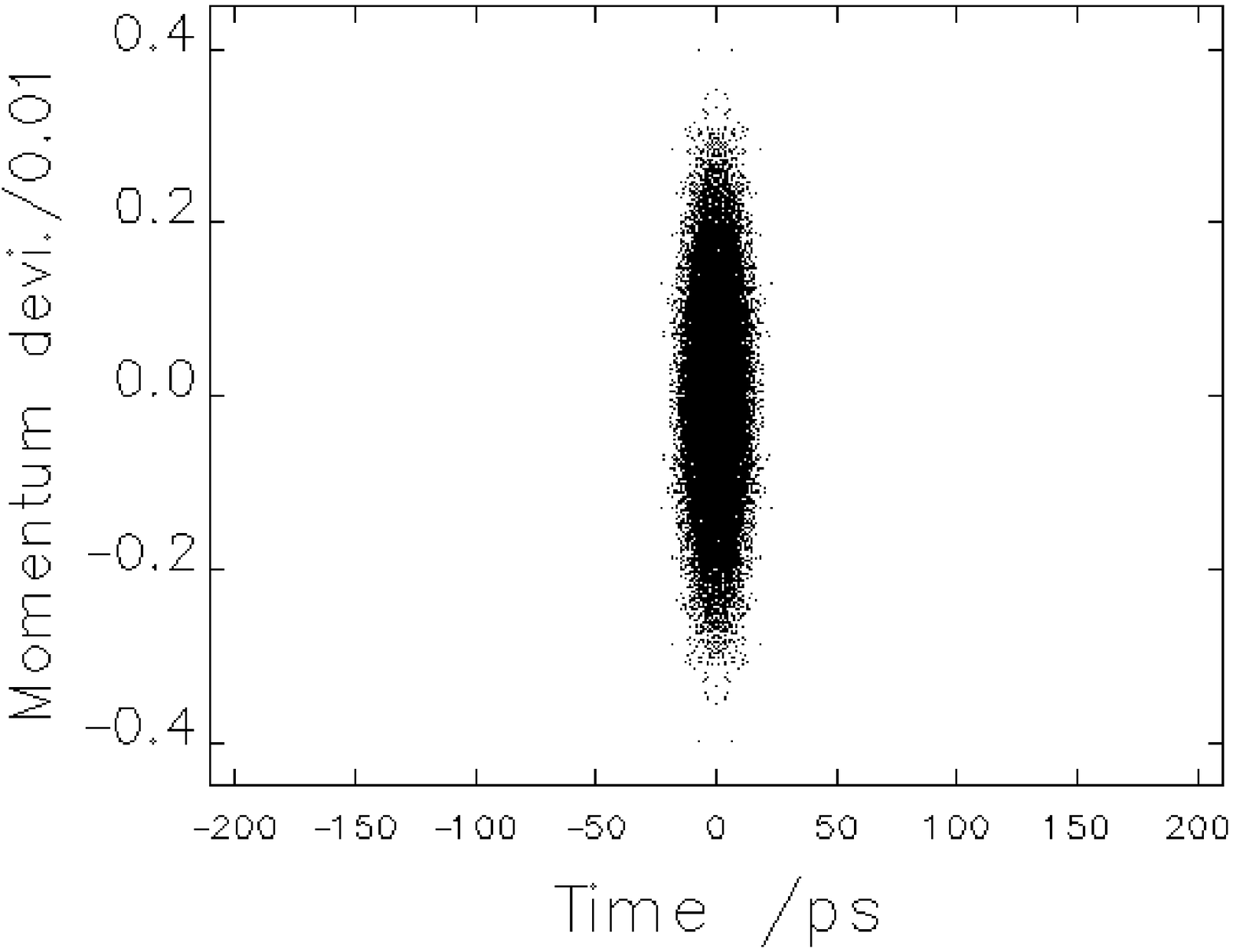} \\ 
  \caption{\label{fig:equilibrium length} Results of longitudinal phase space tracking.Vertical axis indicates the momentum deviation of the bunch, and central momentum of the bunch is about 1566, horizontal axis indicates the longitudinal bunch length in picosecond}
\end{figure}

Particle tracking was performed by ELEGANT to simulate the longitudinal phase space of  the new system with two sc-cavities added . The equilibrium lengths of long bunch (top) and  medium (left bottom) and short (right bottom) are shown in Fig.\ref{fig:equilibrium length}.  
 
\begin{figure}[htbp]
  \centering
     \includegraphics[width=0.9\textwidth]{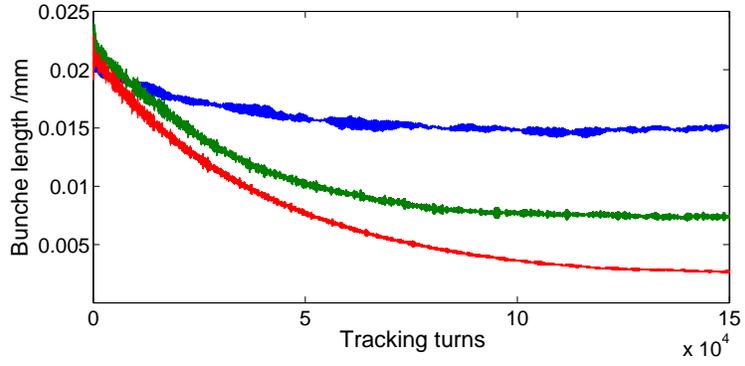} \\
  \caption{\label{fig:length variation of 3 damping} Results of 3 damping times tracking for length variation. Long, medium and short bunches length variation are indicated in blue,green and red. The final results of length after 3 damping times are about 15.1 mm, 7.4 mm and 2.6 mm.}
\end{figure}

A long term of particles tracking of 1000 particles with 3 damping time (about 150000 turns) at the same initial length 20 mm was also performed by ELEGANT, to  simulate the process of  shortening bunches. Fig.\ref{fig:length variation of 3 damping} indicates the length variation of three bunches, where particles start at a long, medium and short bucket location.

The data of 3 damping times can be fitted to get the equilibrium length, and the comparison to theoretical value is given in Table.\ref{tab:fitting length}.
\begin{table}[htbp]
  \centering
  \caption{\label{tab:fitting length} The comparison of fitting and theoretical equilibrium length}
  \begin{tabular}{llll}
  \hline
  &Theoretical & Fitting length & Error \\
 \hline
 Long & 14.8 mm & 14.7 mm & 0.7$\%$ \\
 Medium  & 6.73 mm & 6.59 mm & 2.1$\%$ \\
 Short  & 1.74 mm& 1.69 mm  & 2.8$\%$\\
  \hline
  \end{tabular}
\end{table}

\section{Current Limit}
The increase in the bunch length and energy spread is obvious when beam current exceeds the threshold current \cite{wiedemann2003particle}. To avoid the bunch lengthening, calculation about current threshold is important.

For an electron beam of energy $E_{0}$, a relative energy spread $\sigma_{\delta}$, the beam is described by the longitudinal distribution function $\phi$, which is as a sum of equilibrium distribution function $\phi_{0}$ and a perturbation $\phi_{1} = \widehat{\phi}_{1}\exp[-i \omega s/c+i k z]$, here $k$ is the wave number of the perturbation. And from one-dimensional Vlasov equation for the distribution function\cite{AlexChao}, one can derive
\begin{equation}
1=\frac{i r_{0} c Z(k)}{\gamma}  \int \frac{d \delta (d \phi_{0}/ d \delta)}{\omega +c k \alpha \delta}
\end{equation}
here $Z(k)$ is the impedance,$\delta$ is the relative energy offset of a particle and Gaussian distribution can be used for $\phi_{0}$.

Consider the CSR wakefield generated by an electron moving on a circular orbit of $\rho$ in the middle of two parallel metal plates separated by a distance $2h$,  the shielding parameter is given here \cite{Banethreshold}
\begin{equation}
\Pi=\sigma_{s} \rho^{1/2} / h^{3/2}
\end{equation}
here $\sigma_{s}$ is the bunch length. The longitudinal distribution varies at different bunch length.

For a long bunch,whose shielding parameter $\Pi > 3$, the CSR impedance with  shielding  was calculated by Warnock\cite{csrImshielding}, and analysis of Eq.(6) was carried out for various values of a scaled current\cite{YunhaiTheory}
\begin{equation}
S=2 \sqrt{2\pi} I h/\alpha \gamma {\sigma_{\delta}}^2 I_{A} \sigma_{s}
\end{equation}
where $\sigma_{\delta}$ is the rms relative energy spread, $I_{A}=17045$ A is Alfven current. The beams is unstable when $ S > 6/ \pi$

In HLS II, shielding parameter $\Pi > 3$ corresponds the bunch length $\sigma_{s} > $ 5.8 mm, and both medium and long bunches are in the area. For  bunches whose length is larger than 5.8 mm in HLS II, the current threshold  is
\begin{equation}
I_{\texttt{th}}=I_{\alpha \sigma} \alpha \sigma_{s}
\end{equation}
Here
\begin{equation}
I_{\alpha \sigma}=\frac{3\sqrt{2} \gamma \sigma_{\delta}^{2} I_{A}}{2\pi^{3/2} h}
\end{equation}
here $ I_{\texttt{th}}$ is the threshold of current,  beam becomes unstable when the current is above $I_{th}$. The long bunches current limit is easy to be predicted for the threshold is proportional to momentum compaction factor and bunches length. The physic meaning of $I_{\alpha \sigma}$ could be thought as the current per momentum compaction factor per bunch length , $I_{\alpha \sigma}$ is about $ 112 \texttt{mA} / \texttt{mm} $ in HLS II. As a result, the threshold current of long and medium bunches are about 34.0 mA and 15.5 mA , which are both in large value.

When a bunch is short, whose length is smaller than 5.8 mm in HLS II, the longitudinal wakefield requires special care in the ring for the equilibrium becomes a Haissinski distribution\cite{Haissinski}. The bunch beam theory should be applied \cite{YunhaiTheory}, and the current threshold should be amended as
\begin{equation}
I_{\texttt{th}}=\frac{8 {\pi}^{2} \, (0.5+0.12\Pi) \, \sigma_{s}^{7/3}[fv]_{\texttt{focusing}} f_{rev}}{c^{2} Z_0 \rho^{1/3}}
\end{equation}
here $ Z_{0}=120\pi \,\Omega$, which is the impedance of free space. Bessy II also provides a empirical equation for bursting threshold calculation \cite{Bessymearsure}, and it agrees well with the Eq.(10). So the current threshold of short bunches is about 6.6 mA. 

In comparison with low $\alpha$, using supercongducting RF cavities (SRF) to get short bunches could store more current in buckets. If with low-$\alpha$ optics to shorten the bunch at the same length above (1.74 mm), the threshold is just about 0.1 mA. To make it more clearly, we assume the threshold is $I_{0}=10$ mA at bunch length  $\sigma_{z0}=6$ mm. Then the current limit difference between two methods below 6 mm is shown in Fig.\ref{tab:fitting length}.
\begin{figure}[htbp]
  \centering
  \includegraphics[width=0.8\textwidth]{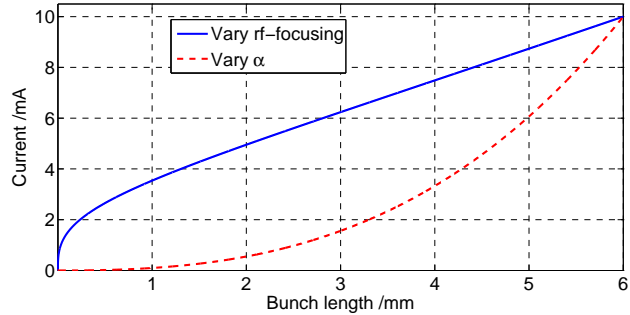}\\
  \caption{\label{fig:two methods current limit} Current threshold when increase rf-focusing and decrease $\alpha$}
\end{figure}

Combined with Low-$\alpha$, bunches can be even shorter. For the new superconducting system above, the momentum compaction factor is about $\alpha$=0.0205, with $\alpha$ to be adjusted to 0.000205, all bunches are expected to be even shorter by 10 times. Finally, short bunches can be about 0.174 mm (about 0.6 ps), and the total current is about 2 mA. The current is small, but it is still much better than Low-$\alpha$ mode. However, for users, a suitable operated $\alpha$ could be chosen for good current and short bunches length at the same time. Table.\ref{tab:choices of alpha} shows typical examples of expected bunches length and total current limit.
\begin{table}[htbp]
  \centering
  \caption{\label{tab:choices of alpha} The short bunch length and total current limit  relation}
  \begin{tabular}{lll}
  \hline
  $\alpha$  &  $\sigma_{s}$ /mm  &  Total current /mA \\
  \hline
   0.0205 & 1.74   &  $>$ 300  \\
   0.0068 & 1.00  & 166  \\
   0.0024  & 0.6   &   41  \\
   0.0017  &  0.4   &   15  \\
   0.0002  &  0.174  &  2  \\
  \hline
  \end{tabular}
\end{table}

\section{Conclusion}
A scheme is presented to operate the HLS II ring with simultaneous medium, long and short bunches. The short bunches current limit can be increased more than 60 times compared to low-$\alpha$ mode. The frequencies of two superconducting cavities  are both near 1.3 GHz,  which are easy to achieve by tuning, for 1.3 GHz is a common frequency for sc-cavities \cite{RFcavitiesbook}. And the voltages are about 2 MV,  which is easy to be realized by single-cell or two-cell structure, so the HOM problems will be simplified. 

\section{Acknowlegements}
We thank A. Chao and J. Wu for useful discussions and encouragements. This work was supported by National Natural Science Foundation of
China under contract No.11327902, No.11175180, No.11175182 and U.S. DOE under Contract No.DE-AC02-76SF00515.

\end{document}